# Flexible Ultrastrong 100-nm Polyethylene Membranes with Polygonal Pore Structures

*Jin LI*[1], Runlai LI*[1], Qiao GU[1], Qinghua ZHANG[1], T.X. YU[2], and Ping GAO[1]†*

[1]Department of Chemical and Biological Engineering, [2]Department of Mechanical and Aerospace Engineering, The Hong Kong University of Science and Technology, Clear Water Bay, Kowloon, Hong Kong SAR, PRC

**SHORT TITLE**

Tough and Flexible 100-nm Fibrous PE Films

**ONE-SENTENCE SUMMARY**

An ultrastrong 100-nm UHMWPE membrane with stretchable interconnected polygonal pores is created, coated with monolayer graphene, and tested for use as a fully conformable piezoresistive skin sensor.

**KEYWORDS**

* Equal Contributions.

†Corresponding Author. Email: kepgao@ust.hk

**ABSTRACT**

Robust nanoporous polymer films with approximate 100-nm thicknesses would be broadly applicable in technological areas such as flexible sensors, artificial skins, separators, antireflection and self-cleaning films. However, the creation of these films has been extremely challenging. To date, all reported ultrathin films are of insufficient mechanical strength for use without substrate supports. We describe here the fabrication of a new ultrastrong and highly flexible ultrahigh molecular weight polyethylene (UHMWPE) 100-nm porous membrane. Scanning and transmission electron microscopy evaluations of film microstructures reveal a planar fibrous structure with randomly oriented polygonal interconnected pores. Moreover, the tensile strength of this film is twice as strong as that of solid stainless steel, with a measured in-plane tensile strength and ductility of approximately 900 MPa and 26%, respectively. We further coated our newly developed film with a monolayer graphene to invent an optically transparent and fully conformable piezoresistive skin sensor and thus demonstrated a potential application.

**MAIN TEXT**

**1. Introduction**

Low-cost, flexible, stretchable, and porous ultrathin polymer membranes are desirable for use in many technologically important applications, including conformable skin strain sensors (*1*), artificial skins (*2*), separators for flexible ultrathin batteries (*3*), filters (*4*), and omni-directional antireflection films (*5*). Such membranes would also be fundamentally important for the elucidation of chain confinement dynamics in ultrathin nanoporous spaces (*6, 7*). Mathematically, the term ultrathin film refers to a homogeneous solid film contained between two parallel planes; this may extend infinitely in the planar dimensions but is restricted in the third (i.e., thickness) dimension to less than 100 nm (*8*). Unfortunately, all previously reported ultrathin polymer porous and non-porous films have yielded insufficient mechanical strengths for use in self-supporting applications.

Currently, ultrathin porous polymer membranes are synthesized using methods such as selective component removal after multicomponent spin-coating (*9, 10*), electrospinning (*11, 12*), and chemical vapor deposition (CVD) (*13, 14*). However, the lack of long-range molecular anisotropy in the ultrathin films synthesized using these technologies renders the films fragile and applicable for use only with substrate supports (*15*). Therefore, alternative methods that can endow high levels of in-plane molecular anisotropy are needed to synthesize freestanding and robust ultrathin polymer membranes.

Ultrahigh molecular weight polyethylene (UHMWPE), which has an average molecular weight exceeding $10^6$ kg/kmol, is considered a gold-standard material for use as a liner of tibial plateaus and acetabular cups in total knee and hip joints, given its robustness, self-lubricating surface, good resistance to creep and chemical attack, and good biocompatibility (*16–19*). High-performance UHMWPE fibers, such as Dyneema$^{TM}$, comprise the strongest and lightest synthetic materials used in the fabrication of bullet-proof vests (*20–22*). The all-planar zig-zag chain packing in the crystal unit cell structure is the theoretical foundation that led to the technological invention of high-performance gel-spun UHMWPE fibers. Based on the orthorhombic unit cell structure and comparisons with the unit cell dimensions of a diamond, Frank estimated that the theoretical tensile modulus of a PE crystal is 250 GPa (*23–25*). Previously, we extended this uniaxial gel-spinning technology by stretching the initially low-entangled UHMWPE in biaxial directions to develop UHMWPE thin film separators for safer rechargeable batteries. The resulting film exhibited a tensile strength of 200 MPa at a volumetric porosity of 76% (*26*).

Graphene, a two-dimensional (2D) single atomic-thin layer of sp$^2$ carbon honeycomb lattice, has attracted enormous research interest because of its high carrier mobility (*27*), mechanical strength and flexibility (*28–30*), optical transparency, and extraordinary chemical resistance (*31, 32*). However, a single layer of atomically thin graphene alone would be too weak for use in applications. Still, the semi-infinite dimensionality of graphene makes it an ideal mechanical filler for providing maximum reinforcement to composites. By designing an alternately stacking and folding method of laminating CVD monolayer graphene onto multilayer polycarbonate ultrathin films, Liu *et al.* prepared a graphene-reinforced composite with an impressive 30% increase in stiffness at a graphene loading of 0.185 vol% (*33*). This suggests that the lamination of graphene onto superstrong ultrathin polymer films would allow the development of fully conformable skin sensors.

This paper aims to report a scalable method for the fabrication of a new and hand manipulatable ultrastrong and ultrathin (thickness ~100 nm) UHMWPE membrane containing topologically interconnected polygonal pores. This fabrication method utilizes a novel biaxial stretching technique in which initially low-entanglement UHMWPE precursors are stretched biaxially in a semisolid state. Using this method, fibrous films with approximately an in-plane tensile strength of 900 MPa and ductility of 26% were generated. To demonstrate the potential use of this film as a flexible and transparent gesture sensor, we invented a conductive piezoresistive sensor by coating the UHMWPE film with a monolayer of CVD graphene. For the first time, therefore, we have developed an ultrathin flexible piezoresistive sensor with a high gauge factor of 74 at 5% strain. We then mounted this piezoresistive sensor to a wrist using van der Waals (vdW) interactions and observed a remarkably high level of sensitivity to the flexion and extension of this joint.

2. **Results**

**2.1 Macroscopic Properties of the 100-nm UHMWPE film**

Macroscopically, the newly fabricated ultrathin UHMWPE porous films were mechanically robust and ductile, optically transparent, and chemically hydrophobic to water but organophilic to oil. Before demonstrating these macroscopic features, we first confirmed the thickness of the film by stylus profilometry. Subsequently, the macroscopic features were assessed using a combination of photographs, videos, and ultraviolet-visible (UV-vis) spectroscopy characterizations. The macroscopic properties of the film were qualitatively characterized by subjecting the film to experiments involving alternating water and acetone jet spattering in the outdoor sportsground of the HKUST campus. Briefly, we sprayed both sides of the film surface by alternately squeezing pure jets of water and acetone from narrow mouth nozzles attached to squeeze wash bottles. Quantitatively, the film was tested on a customized indentation testing stage in conjunction with the Advanced Rheometric Expansion System (ARES). Square-shaped samples with surface areas of 65 mm × 65 mm were confined inside carbon fiber frames for use in these macroscopic characterization experiments. A photograph depicting the experimental setup containing the film sample is displayed in **Figure 1A**. Here, a friction-free 15.7-mm-diameter glass ball was used to impose an indentation load on the film. This diameter was selected to limit indentation stress and avoid highly localized deformation (*34*). The colorful circular stripes on the photograph were the interference patterns between the deforming film and the floor-ceiling lights in the laboratory.

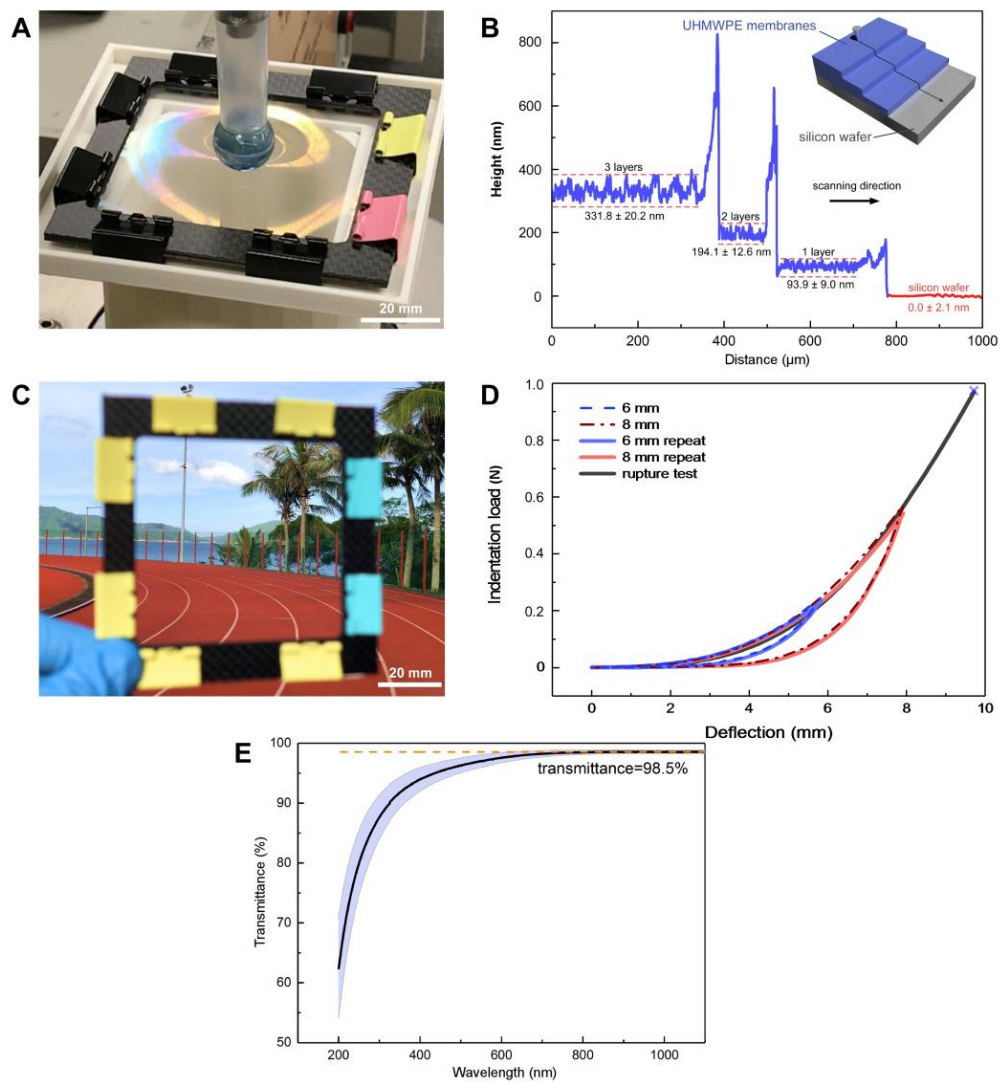

**Figure 1.** Images and plots of the thickness, indentation responses and optical transmittance of the UHMWPE film. (A) A photograph of the UHMWPE film sample on a 3D printed fixture during indentation load-deflection test under ARES-2000. The colorful strips in the photo represent the interferences between the deforming film and the floor ceiling lights at a deflection of 1.25 mm. (B) Plots of film heights versus distance along the stepwise stacks on a silicon wafer surface measured by stylus profilometry. The inset on the top right corner depicts a schematic illustration of the stacking configuration. (C) A photograph of the UHMWPE film confined inside a carbon-fiber frame taken in the HKUST's outdoor sportsground. (D) Indentation load versus deflection responses of the UHMWPE film measured at 25 mm/min at room temperature. Two repeated loop tests at limiting deflections of 6 mm (dashed blue and continuous light blue) and 8 mm (long dashed and dotted red and continuous light red). The continuous black line is the rupture test. (E) UV-Vis spectra of the UHMWPE film displayed in **Figure 1D** measured over eight different locations along the film surface (**Figure S3A**), four were along the corners, and four were at the middle of each edge of the film. The black line is the mean relative transmittance intensity and the shaded light purple region is the error band. A horizontal dashed orange line at transmittance 98.5% is also shown in the plot.

**Figure 1B** describes the use of stylus profilometry to measure the thickness of the film. Stepwise-stacking method was used for the measurements, and a schematic sketch of the stacking setup is shown as an inset at the top right corner of **Figure 1B**. As shown, the measured film height clearly decreased in a stepwise manner as we moved from the triple-layer region to the bare silicon wafer surface, except at the interfacial boundaries between different film stacks where folding caused rapid local spikes in the measured overall film thickness. The average thicknesses per film layer, in the triple-, double-, and single-layer regions were $110 \pm 7$ nm (total: $332 \pm 21$ nm), $97 \pm 6.5$ nm (total: $194 \pm 13$ nm) and $94 \pm 9$ nm, respectively. Although triple-layer profilometry yielded a higher film thickness when compared with those measured in the double-layer and single-layer regions, these variations were within 15%. Together with the cross-sectional SEM observation, this finding confirmed an approximate film thickness of 100 nm.

**Figure 1C** shows an image of the highly transparent UHMWPE film taken in the HKUST's outdoor sportsground. The qualitative features of the film are shown in the **SI Movie** recording of the impingement response of the film to water and acetone jets (see **SI Movie, Episode I & II** in the supplement). **Episode I** depicts the interactions between the water jets and the film surface, while **Episode II** depicts the impingement of acetone jets on the film from the backside of the membrane surface. One can observe four characteristic features from these movie clips. First, the film is hydrophobic and organophilic. The hydrophobicity is demonstrated clearly by the pinning effect of the water droplets on the film after water spattering, which exhibited optical rings when viewed from the opposite side of the film. The organophilicity is demonstrated by the rapid spreading of acetone on the film upon impingement. Second, the film has a very high level of mechanical stiffness, which is inferred from the clear audible sound produced upon each impact and the fact that the phonon frequencies are proportional to the square root of the mechanical stiffness of the film. Third, the film is strongly viscoelastic, as demonstrated by the rapid disappearance of wrinkles produced by acetone spattering. To demonstrate this effect more clearly, we captured four still images immediately after the impact of an acetone jet from **SI Movie Episode II** and exhibited these images in **Figures A–D** of **Figure S1. Figure S1A** shows a set of hyperbolic wavelike wrinkles formed on the film immediately after acetone impingement. This is due to the energy transfer from the acetone jet to the film surface and the subsequent propagation of energy through the film in the form of out-of-plane oscillations. The film smoothness was restored within 0.12 seconds after impact, implying the film can exhibit self-healing property (see **Figures S1 B-D)**. Fourth, the film is porous, as demonstrated by the unpinning of water droplets after acetone spattering. When the acetone jet impacted the back of the film, the pinned water droplets on the front become unpinned and slid away from the film surface. As acetone was imbibed in the film, it replaced the air trapped under the water droplets, thus removing the pinning force between the water droplets and film. The use of liquid imbibing to displace air in porous films was identified as a useful property in the design of liquid-repelling self-repairing slippery surfaces (SLIPS) (*35*). In the next section, we further investigate the topology of the pores in the film and demonstrate that the pores comprise topologically interconnected polygons with vertices joined by planar nanofibrils that form a cellular nanostructure dominated by two-dimensional stretching. The pore structures were characterized using scanning electron microscopy (SEM), transmission electron microscopy (TEM) and Brunauer–Emmet–Teller (BET) isotherms.

**Figure 1D** depicts the quantitative characterizations of the film robustness regarding the indentation load versus deflection curves. Two different film specimens were subjected to repeated loop tests, and the data are represented by continuous and dashed/and long dashed and dotted lines. These results depict at least two characteristic features. First, the film is strongly viscoelastic, as demonstrated by the congruency of the indentation-displacement curves at different limiting deflections; all indentation loads at the same deflection were nearly identical. Second, the film is highly flexible and robust, with a maximum rupture load and deflection of 97 g and 9.7 mm, respectively. Astonishingly, the maximum deflection with respect to the thickness and surface area of the film is equivalent to the ability of a square-shaped silk scarf with a thickness of 500 μm and a surface area of 300 × 300 m² to deflect vertically by 50 m before rupture. We applied the approximate solution presented by the Föppl–Hencky membrane theory by Begley and Martin regarding a circular indentation of the film: $\frac{c}{R} = \sqrt[4]{\left(\frac{3}{2\pi}\right)\frac{P_0}{EhR}}$, where $c$ is the contact radius between a ball of radius R (15.7 mm) and the film, $P_0$ (0.95 N) is the maximum indentation load, $E$ is the tensile modulus, and $h$ is the film thickness. By substituting the value $E = 8.6$ GPa (see next section), we determined that $c = 4.04$ mm. Accordingly, this 100-nm film could withstand a pressure of 0.18 bar before rupture (*36*).

The newly observed loading/unloading indentation characteristics comprise another intriguing feature of these ultrathin films. The glass ball and film surface remain in full contact during the entire unloading process at a testing speed of 25 mm/min, which is quite high for indentation measurements. This behavior suggests two unique features of the ultrathin films, namely viscoelasticity and a contact surface that is fully conformable to glass surfaces. **SI Movie, Episode III**, provides visual evidence to corroborate this finding. No similar reports of nanofilms are available, and previous reports of needle withdrawal after puncture involved fractures prior to ultimate failure such that the unloading process corresponded to friction between the needles and the fractured interface (*37*).

**Episodes III and IV** in the supplement movie present video recordings of the indentation-deflection responses at 25 fps. To enhance the visual effect, we purposely included an interference pattern produced by the tubular ceiling lights in the laboratory while recording the video, as interference patterns have previously been used to measure deflection strains in ultrathin films (*10*). As the UHMWPE film herein was readily stretchable on a traditional tensile testing setup, we aimed to measure Young's modulus directly, rather than through an indirect solution provided by the idealized Föppl–Hencky membrane theory.

**Episode III** depicts load-deflection loop tests at limiting deflections of 6 mm, together with the real-time load-deflection curve. Clearly, circular interference patterns erupt from the center of the indentation and propagate radially outward as the indentation load increases. Meanwhile, upon reversal, the interference pattern nearly mirrors the indentation process, in which the same deflection load yields the same interference pattern. To further characterize this mechano-optical response, we obtained screen shots of the interference patterns at 5.0 g and indentation load during loading and unloading processes, respectively, and presented these images in **Figure S2A–B**. At the same indentation loads of 5 g, the interference patterns are almost mirror images. Similarly, the interference patterns recover completely even after the 8 mm deflection loop test. In addition, when the film ruptures, we can observe the release of a phenomenal amount of fracture energy through the audible noise released. The large undulations of the ruptured film and the optical interferences generated in these ruptured films allow a closer examination of the ultrafine film thickness, as shown in **Figure S2C**. The noise generated upon film rupture was consistent with the high estimated energy required to cause a fracture. As to be shown in the next section, the film exhibits a high energy to fracture of 166.6 ± 30.8 MPa. The observed mechano-optical correlation here may lead to the development of optical sensors that transduce strains into interference patterns. One practical application of this technology might involve the earlier detection of failures in structural windowpanes, where the optical transmittance is crucial and thus excludes all other sensing devices that might compromise this parameter.

In addition to optical transparence in the visible region, as demonstrated in the above photos and videos, the high bandgap of polyethylene, 7.35 eV, indicates that ultrathin UHMWPE films would be optically transparent at wavelengths exceeding 173 nm in the ultraviolet (UV) (*38*). This feature can be illustrated quantitatively by the ultraviolet-visible (UV-Vis) spectrometric data of UHMWPE films, as shown in **Figure 1E**. To assess the optical homogeneity of the film, we conducted measurements at eight different locations on the film; each measurement was conducted on an irradiation area of 1.0 mm × 0.9 mm. The optical transmittance of the film exceeded 95% at wavelengths between 400 and 1100 nm and 98.5% at wavelengths greater than 500 nm, indicating an even better performance than the atomically thin monolayer graphene. The minimum optical transmittance was 60% at 200 nm. We speculate the decrease in optical transparencies at wavelengths below 400 nm is the result of the combined effect of interfacial scatterings in the amorphous/crystalline interface of the polymer substrate and a tunneling effect as the wavelength approaches the bandgap region of the semiconductor film (*39, 40*). We additionally used the Lambert-Beer law to calculate film thicknesses using the UV absorption intensity at wavelengths of 200, 250, and 350 nm. Here, we selected eight irradiation areas in the film for thickness measurements across the film surface, see **Figure S3A** for the specific location of the irradiation area. **Figure S3B** shows the corresponding thickness values measured over a number of different film specimens with different nominal thickness. Note that the results obtained at all three wavelengths agree perfectly with the Lambert law, and the calculated thicknesses agree extremely well with the stylus profilometry measurements presented in **Figure 1B**. Therefore, all mechanical data presented below are based on the thickness values calculated in **Figure S3B**.

## 2.2. Nanostructure and Tensile Properties

We investigated the nanostructure of the UHMWPE film using SEM and TEM. For the SEM surface characterization, we coated the UHMWPE film onto a conductive silicon wafer surface covered with a 3-nm platinum film. This is for two reasons. Firstly, because the film is highly porous and is extremely thin, current tunneling from the platinum on the silicon surface is sufficient to minimize surface charging effect. Secondly, conductive sputtering on UHMWPE surface directly can cause structural damages and artifacts in structural dimensions, particularly to the fine fibrils in the film texture (*41*). To characterize the thickness, we sandwiched the sample between two Nafion films with approximate thicknesses of 1 μm and fractured the composite film in liquid nitrogen. Details of the Nafion loading protocol are given in a separate manuscript that describes the use of the composite membrane as an ion-exchange membrane for fuel cells and redox flow batteries (*42*). (Unpublished result).

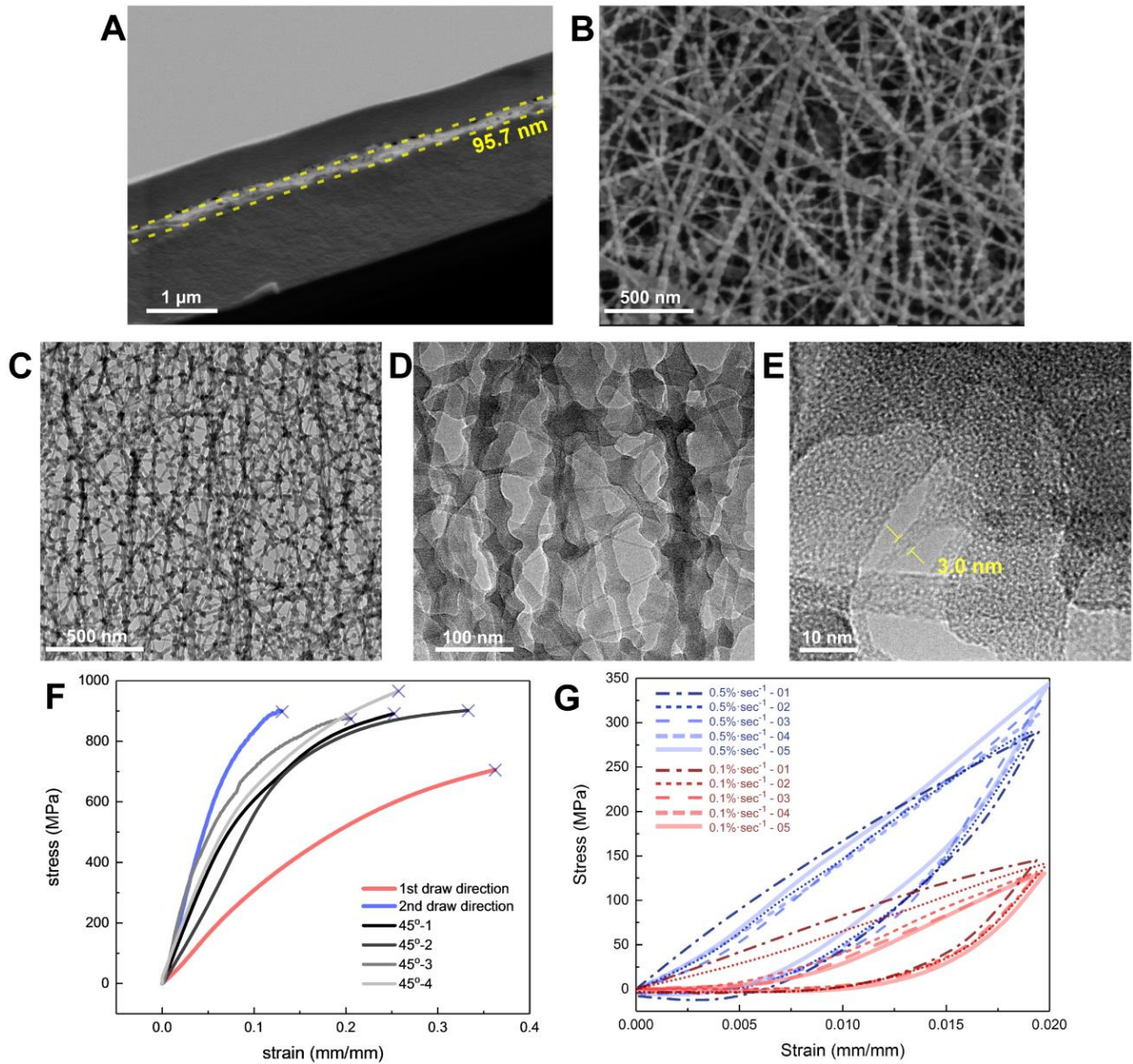

**Figure 2.** Nanostructure and tensile properties of the UHMWPE film. (A) SEM of the cryogenically fractured cross-section of the Nafion/UHMWPE/Nafion film. The two dotted yellow lines inscribed in the image were drawn along the interfaces between the UHMWPE and the Nafion film. The measured gap between the two dotted yellow line is 95.7 nm. (B) SEM micrograph of the film's surface topology. (C-E) TEM micrographs of the UHMWPE films at different magnifications. The smallest nanofibril in the UHMWPE film is ~3 nm as indicated in Figure 2E. (F) Tensile stress-strain curves of UHMWPE films measured at the ambient condition at $0.1\% \ s^{-1}$. The six curves were obtained on specimens tested along the second draw direction (blue), first draw direction (red) and at 45 ° to the draw directions (different levels of gray, four repeated specimens). (G) Tensile loop stress-strain curves tested at two different Hencky strain rates of $0.1\% \ s^{-1}$ (red) and $0.5\% \ s^{-1}$ (blue), respectively, at limiting extensional strain of 2% and at room temperature. Five loops were conducted at each strain rate and the sequence of the loop is labelled sequentially by the last two-digit-numerals at the top left corner of the plot legend.

An SEM micrograph of a cross-section of the Nafion/UHMWPE/Nafion film is depicted in **Figure 2A**. The darker regions represent cross-sections of the Nafion films and the brighter cross-section of the UHMWPE film. The strong color contrast is caused by the density differences between these two polymer films. The density of the Nafion film was 1.95 g cm$^{-3}$, which is at least twice the density of PE. The high porosity of the UHMWPE film further reduced the density while increasing the phase contrast. By inscribing two dotted yellow-colored parallel lines along the Nafion/UHMWPE interface, we obtained an average UHMWPE film thickness of 95.7 nm. This morphological examination further validated our thickness measurements determined via profilometry. We may therefore conclude confidently that the newly prepared UHMWPE film thickness is approximately 100 nm.

An SEM micrograph depicting the surface topology of the film is shown in **Figure 2B**. The overall film structure comprising randomly oriented fibrils and structural homogeneity is shown in the lower magnification micrograph in **Figure S4**. We may discern three characteristic features from **Figure 2B** and **Figure S4B**:

(1) First, the film comprised nanofibrils that form interconnected polygonal pores with stiff edges and are randomly oriented in the plane of the film. Most fibrils had approximate widths of 20 nm, although a few had approximate widths of 40 nm. These thicker fibrils appear to comprise multiple thinner fibrils bundled together by edgewise attachments. Furthermore, by focusing on the profiles of the fibrils presented in **Figure 2B** (see also the lower magnification image in **Figure S4B**), we could discern additional topological features. The wider fibrils running through the plane of the film overlapped to form principally 4-edged polygons, while the thinner fibrils within the 4-edged polygons formed the struts of triangular-shaped polygons that helped to stabilize the structures. These structures satisfy the Maxwell stability criterion for stretch-dominated cellular structures (*43*).

(2) Second, fibrils with widths greater than 10 nm exhibited a "shish-kebab" morphology, wherein epitaxially grown kebabs known as folded chain crystals were formed by a thermal annealing process after film stretching at high temperatures (*44*).

(3) Thirdly, no lamellar overgrowths (or kebabs) were discernable on fibrils with widths less than 10 nm. Further structural corroborations were provided by the TEM observations, as shown in **Figures 2C–E**.

**Figure 2C** depicts the overall film texture under bright-field TEM. The contrast in this image resulted from electron scatterings produced by the polymer substrate; the darker regions indicate polymer fibrils, while the bright background indicates the voidage in the sample. In addition to the topological features observed by SEM, the micrograph in **Figure 2C** depicts further topological features normal to the plane of the film. We can readily observe that the interconnected polygonal pores are stacked into a multilayered structure along the thickness of the film and that the polygons are randomly oriented within each layer. In addition, the fibril width distribution is significantly narrower. We could ascertain up to 4 layers of fibril-stacks with topological features that appeared to be invariant between layers. It is also clear that the fibrils with widths less than 10 nm do not contain epitaxial crystal overgrowth. In addition, the narrowest fibrils in the UHMWPE film are about 3-nm in width as inscribed in **Figure 2E**. The 3-nm fibrils are also discernable from the lower magnification image in **Figure 2D** in which these long and thin fibrils are apparently elongated fibrils that act to bridge the wider shish-kebabs. To our knowledge, this is the lowest fiber dimension ever observed in PE structures. Up to now, it had been accepted that the smallest PE fibril diameter in oriented PE structures was about 15 nm due to chain confinement effects (*45*). While it is not clear why PE fibrils in the ultrathin films are so different from those in the uni-directionally drawn fibers, the discovery here may lead to further breakthroughs in elucidation of deformation mechanisms.

The term shish-kebab was first coined by Keller in 1974, following the first observation of the unique PE crystals formed from sheared solutions by Pennings in 1965 (*44, 46*). The discovery of this structure led to a major technological breakthrough in the creation of the toughest and lightest ballistic-proof UHMWPE fibers in the world (*47*). Since then, similar structures have been reported in numerous other systems (*48, 49*). As the aim of this communication is to announce the discovery of our new ultrastrong nanofilm, we will continue these structural elucidations in a separate publication.

Individual fibrils of different widths (from 3 nm to 20 nm) in the above TEM micrographs all seem to depict similar gray levels, suggesting that the fibrils of different widths are of similar thicknesses and therefore are of nano-ribbon morphology. To confirm this suggestion, we performed a systematic image analysis of **Figures 2D** and present the image analysis data in **Figure S5** in the supplement. Essentially, the electron scattering intensity is independent of the fiber width, and the intensities of two overlapping fibrils are twice the value of an individual fibril. Interestingly, the kebabs also appear to exhibit similar planar configurations. BET isotherm measurements provide additional evidence for the planar structure. The nitrogen absorption/desorption isotherms depicted in **Figure S6** show that the film exhibited Type IV isotherms with planar or H3-type pores (*50*).

**SI Movie, Episode V** provides a more vivid demonstration of the fibrous porous structure and thinness of our new UHMWPE film under TEM. Here, we made *in situ* video recordings of the film edges on the copper grid manipulated by charges produced by electron beam irradiation at an accelerating voltage of 200 kV. The electrical charges on the folded edges of the UHMWPE film moved the fibrous film up and down in repetition, like the mechanical opening and closing of a book. During this movement, we can also acquire a glimpse of the film thickness.

**Figure 2F** presents the ultimate tensile performance of the film in terms of the engineering tensile stress-strain curves measured at a constant Hencky strain rate of $0.1\% \, s^{-1}$ and ambient temperature. Variations in film thickness were characterized using UV-vis spectroscopy, which yielded an average film thickness of 95.8 ± 14.3 nm. Six specimens were tested: two parallel to the two different draw directions in the film, and four along the diagonal (45° to the two draw directions). The sample dimensions used for the tensile tests had respective widths and lengths of 5 and 8 mm. The UV-Vis testing result and the specimen configurations are given in **Figure S3A–B** in the supplement. We attribute the high tensile properties of the film to the orientation factors of molecular anisotropy and interconnected nanofibrous polygonal pore structure, as revealed by both SEM and TEM. The strong molecular orientation produced using our novel method effectively aligned the polymer chains in the plane of the film, thus enabling the formation of highly anisotropic planar fibrous ribbon structures. These ribbons formed the struts of the interconnected polygonal pores, thus satisfying the Maxwell stability criterion for a stretch-dominated two-dimensional topological structure with a low level of stiffness but high tensile strength (*51*).

For the first time, we have created 100-nm-thick porous films with tensile strengths of ~900 MPa (912 ± 35 MPa), a Young's modulus of ~8 GPa (8.6 ± 3.1 GPa), and a ductility of 26% (0.26 ± 0.05 mm/mm). Furthermore, no apparent yielding was observed on the engineering stress strain curve, even at the high tensile strain of 36%. The ultimate tensile strength of the film is twice that of solid stainless steel. Previously, the highest reported tensile strength in the literature referred to a spider fiber ultrathin film by the self-reinforced *B. mori* silk fibroin (*10*), which had an ultimate tensile strength of 100 MPa. However, the brittleness of the spider film rendered direct measurements impossible, and indirect measurements such as puncture pressure and interference patterns on supported films were used to determine the stress-strain relationship. The authors attributed the brittleness in the ultrathin film to the chain confinement effects in spider fibrils. To our knowledge, no previous report has discussed the direct tensile testing of ultrathin films.

**Figure 2G** depicts the viscoelastic properties of the film at two different Hencky strain rates ($0.1\ \%\ s^{-1}$ (orange) and $0.5\ \%\ s^{-1}$ (blue)) measured at $45°$ to the orthogonal draw directions. The corresponding average values of Young's moduli were $8.6 \pm 1.26$ GPa and $16.4 \pm 1.13$ GPa, respectively, thus demonstrating that the film at $0.5\ \%\ s^{-1}$ is nearly twice as stiff as that at $0.1\ \%\ s^{-1}$. The latter is consistent with the data presented in **Figure 2F**, although the standard deviation is lower. Further, at both strain rates, the loop tests were nearly congruent after the initial two loops (long dash-dotted line and dotted lines), suggesting that the film has strong linear viscoelasticity. Finally, the film's ultimate elastic modulus can be inferred from the initial unloading slope. The measured Young's moduli (by taking the maximum slope of the unloading curves) at the two different strain rates were $35.6 \pm 2.6$ GPa and $29.1 \pm 1.9$ GPa, respectively. While the moduli values differ by approximately 18% using the unloading curves, the difference was significantly smaller than the 47.5% variation calculated using the loading curves at different strain rates. This result further demonstrates that the loading process is dominated by viscoelasticity, whereas the unloading is elastic.

**2.3. Application of the Ultrathin UHMWPE film as a Piezoresistive Flex Sensor**

Having created the ultrathin UHMWPE film, we herein describe one application of this new film to illustrate how it could facilitate the creation of unique and previously unimagined devices. UHMWPE represents a gold standard material in artificial total replacement joints, given its proven usefulness and good biocompatibility (*19*). We invented a conductive piezoresistive flex sensor by coating this new UHMWPE film with a CVD-grown graphene monolayer. The details of the device and structural-property characterizations of the composite film are a topic for a separation publication (*52*). We first affixed two 9-µm-thick copper foil electrodes with surface areas of 2 mm$^2$ to the film surface using conductive silver paint. We then mounted two 50-µm-diameter copper wires to the copper foil electrodes using silver paint. The surface area of the flex sensor was approximately 10 cm$^2$, and the two electrodes were placed at an interval of 3-cm in the middle of the strip. For quantitative piezoresistance measurements using the ARES, rectangular films with surface areas of 10 mm × 15 mm were used, and two electrodes were placed at a 1-cm distance along the stretching direction. The resistance versus wrist movement, as well as the uniaxial tensile loop strain, were measured using an ohmmeter. Simultaneous load-displacement and resistance-displacement curves are measured at a constant Hencky strain rate of 0.1 % $s^{-1}$ at the ambient conditions. We used both photos, videos, and quantitative measurements to demonstrate the performance of this new device.

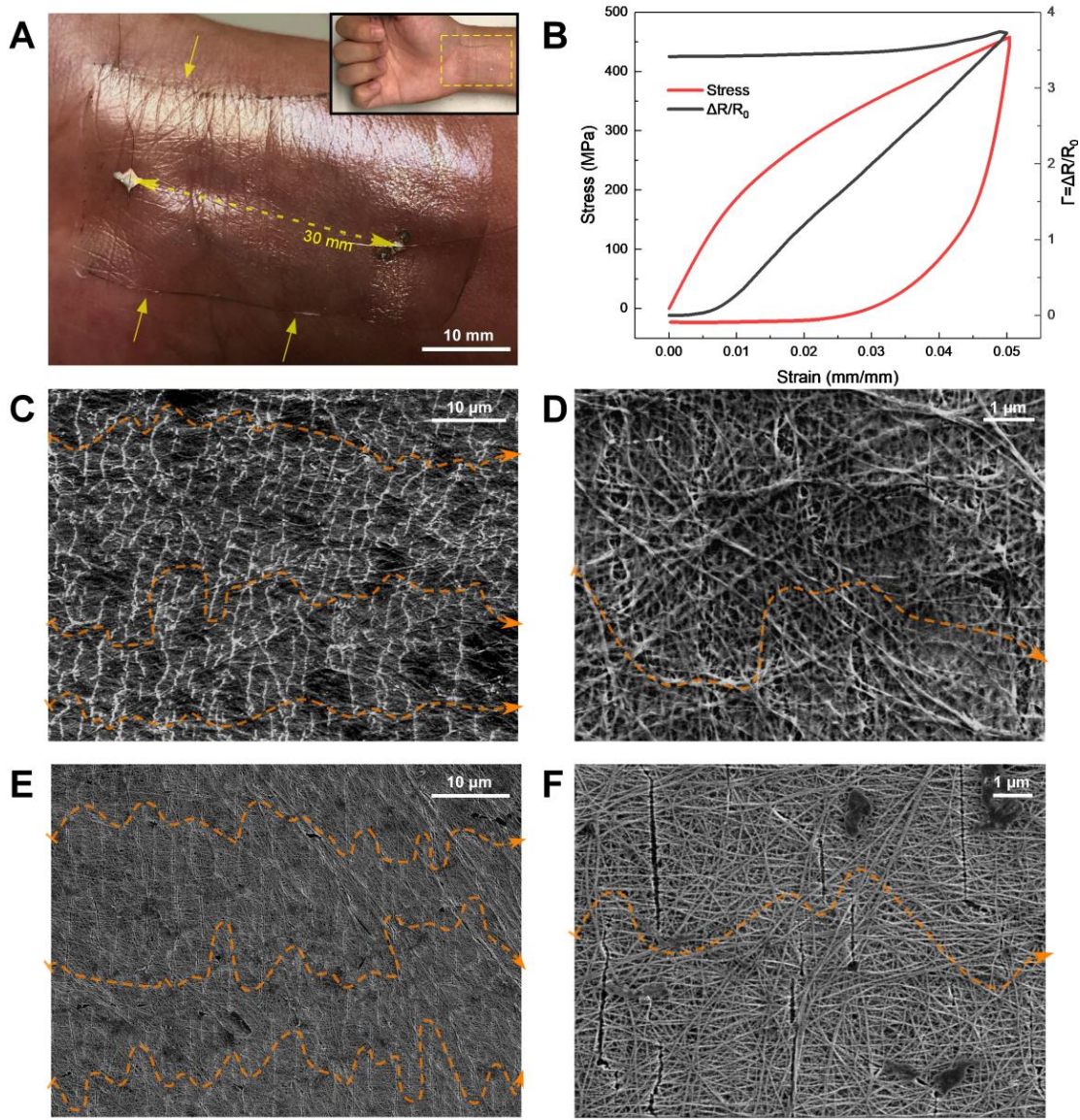

**Figure 3.** Images and characterization of graphene/UHMWPE flex sensor. (A) Photograph of the conductive film mounted on an arm wrist. Two copper foil electrodes of 4 mm$^2$ were attached to the film surface at a separation of 30 mm (left-right dashed arrow in yellow color) by conductive silver paste. Film folding along the edges are labeled in yellow arrows inscribed in the image. 50 μm-diameter copper wires were connected between these electrodes and an ohmmeter for measuring resistance. The insert on the top right corner shows the mounting location of the film on the wrist and highlighted in dashed yellow rectangle. (B) Simultaneous tensile stress-strain (red) and piezoresistance-strain (black) loop test of the film at 5% limiting strain tested at 0.1 % $s^{-1}$ at room temperature. (C-D) SEM micrographs of the sensor surface after unmounting from the wrist at two different magnifications. (E-F) SEM micrograph of the simulated sensor after 2% strain. Three orange colored rightwards dashed arrows from bar were inscribed in Figures 3C and 3E, and one in Figures 3D and 3F. All arrow lines were drawn through the visibly intact graphene surfaces along the stretching direction of the film.

**Figure 3A** presents a photograph of the flex sensor mounted on arm wrist. The insert on the top right corner shows the location of the sensor on the wrist joint. The full skin conformability of the film is clearly demonstrated by the fine skin texture, which is sharpened by the shiny graphene coating. The slight grayish color is due to light absorption by the graphene monolayer (*31*). Conformability is visible everywhere except at the edges of the film, where folding of the graphene-coated film caused darker edges and indicated by yellow arrows. A video of the flexion and extension of the wrist and the corresponding ohmic resistance responses is presented in **SI Movie, Episode VI**. A high gauge factor can be inferred from the rapid changes in the resistance as the wrist joint undergoes flexion and extension processes. The resistance between the two electrodes doubled from 13 to 26 k$\Omega$ as the unclenched fist, with an initial flexion angle of 77º, slowly returned to the neutral position. Upon wrist extension by 86º, the resistance increased sharply to 34.28 k$\Omega$. During the reverse process, the ohmic resistance returned to 15 k$\Omega$ at a flexion angle of 81º. Although hysteresis was apparent in the flexion/extension process, this hysteresis was substantially lower than that observed in the uniaxial extension/compression of the film (see **Figure 3B**). The smaller hysteresis shown here is attributed to the lateral constraint caused by skin/polymer bonding, which was absent during the uniaxial stretching process. To demonstrate the robustness of the film, we also recorded the unmounting process in **Episode VII**. Specifically, we placed two lengths of clear adhesive tape on the two ends of the flex sensor during the unmounting process to minimize wrinkles in the film upon detachment from the skin. The entire unmounting process, depicted in **Episode VII**, not only allows one to witness the robustness of the film but also shows how easy the film can be manipulated manually. One immediate implication from this experiment is the potential use of this film as a breathable bandage for large-area wounds, during which the film could also sense the cell regeneration process. Moreover, autologous skin cells and/or drugs could potentially be preloaded inside the pores of the membrane to allow more rapid skin regeneration.

**Figure 3B** depicts the quantitative piezoresistance sensitivity test in which the left y-axis represents the nominal tensile stress and the right the resistance ratio, Γ. Here, Γ is defined as the change in the ohmic resistance, $\Delta R$, divided by the initial resistance, $R_0$, such that $\Gamma = \Delta R/R_0$. The red curve represents the viscoelastic stress-strain loop test during the loading and unloading process, while the black curve indicates the corresponding resistance ratio. The gauge factor, $GF$, was used to quantify the sensitivity of the piezoresistive sensor. Using the definition of $GF$ = $\Gamma_{max}/\varepsilon_{max}$, a value of 74 was calculated at an applied limiting extensional strain of 5%. This is the highest value ever reported using piezoresistive sensors prepared by monolayer graphene (*53, 54*). **Figure 3B** also depicts the linear dependence of piezoresistance under increasing extensional strain. Note that a smaller gauge factor is observed at strains less than 0.4%. However, the resistance decreases significantly more slowly upon unloading than during the extension process. Here, Γ depicts a nonlinear dependence again strain to yield a small gauge factor of 6.8. The undesired hysteresis is also observed in other piezoresistive sensors (*55*). However, film sensors for wounds maybe deposable given the fact that the 100-nm sensor might be needed for detailed spectroscopic and chemical analysis after detachment. On the other hand, the flexion sensor on the skin exhibited a much smaller hysteresis, which suggests the lack of a lateral constraint might have been a limiting factor affecting this irreversibility. In fact, the higher reversibility of the indentation-deflection loops depicted in **Figure 1D** further corroborates the significance of the lateral constraint in terms of more rapid structural recovery. Future experiments of piezoresistivity can be conducted in conjunction with the indentation-deflection fixture depicted in **Figure 1A** to confirm the effect of lateral constraint. Thus, we surmise the present experiment provides a lower limit of recoverability of the strain gauge factors.

The microstructural changes in the film after the flexion test and in a film that had been stretched by 2% via uniaxial extension are depicted in **Figures 3C–F**. **Figures 3C–D** present SEM micrographs of the surface of the unmounted flexion sensor at magnifications of 1900 × and 13000 ×, respectively. Clearly, after the flexion and stretching tests, nearly parallel white striations that were nearly normal to the wrist could be observed. The higher magnification micrograph in **Figure 3D** shows that these striations correspond to cracks in the monolayer graphene. The average distance between the gaps in the fractured graphene was approximately 150 nm, which is well beyond the distance for current tunneling (*56*). However, detailed analysis on the micrographs suggests that the cracks in the graphene after wrist flexion are not continuous, and by tracing unfractured graphene on the film, we were able to draw at least three continuous lines through the film. This suggests that the change in resistance is mainly caused by the change in charge conduction path. To further verify this hypothesis, we prepared a film that was subjected to a 2% limiting strain after the uniaxial loop test. SEM micrographs of the surface morphology at magnifications of 2000 × and 10000 × are shown in **Figure 3E–F**. The stretching direction was normal to the striations shown in the micrographs. Although longitudinal cracks normal to the stretching direction were again observed, these cracks were narrower and the film was visibly smoother. This discrepancy is attributed to the wrinkles present in the film unmounted from the wrist. One common feature of the tested films was the lack of complete cracks in the graphene surface traversing the film in the normal direction to stretching, which again allowed one to draw a continuous path along the draw direction of the film. These results further corroborate that the increase in resistance is due to an increase in the conduction distance, whereas the overall reduction in graphene conductivity is a result of the cracks in the film **(Figure 3C-F)**.

The newly discovered high-gauge factor in monolayer graphene, as well as the full skin conformability of the newly developed piezoresistive sensor, paves the way for new design concepts regarding mountable skin sensors. Hitherto, mountable skin sensor designs have been guided by the idea of soft gel-like materials that would mimic the low stiffness of skin (*57*). However, such designs inevitably yield films that are too thick to achieve full conformability and breathability. The observed full skin conformability can be attributed to the ultra-low bending modulus of the UHMWPE film, $B = \frac{Et^3}{12(1-\nu^2)}$. Here, $B$ is the bending modulus in $J$, $E$ is the Young's modulus in Pa （~10 GPa）, $t$ is the film thickness in m ($10^{-7}$m), and $\nu$ is the film's Poisson ratio. Using the properties of the film and assuming $\nu = 0.3$, one can estimate the film's bending modulus is $\sim 10^{-12} J$, which compares favorably with the epidermal skin (*58*). Additionally, the high porosity in the UHMWPE film means it is also potentially applicable as a fully conformable sweat sensor, use of which will help to simplify the sweat sensor designs for sweat retention (*59, 60*). Furthermore, we used the low bending modulus of the film to prepare a Turing-patterned hybrid membrane with enhanced ductility and mechanical stiffness (*61*) (unpublished result).

**Discussion and Conclusions**

This paper demonstrates that a novel, robust, mechanically flexible, 100-nm UHMWPE film with polygonal pore structures can be fabricated in a readily scalable manner using a preparation method that involves an initially low entanglement UHMWPE film and subsequent biaxial stretching. By coating this optically transparent film with monolayer graphene, the resulting film exhibited high piezoresistive gauge factors and acted as an effective flexion sensor. Structural investigations demonstrated that the film is essentially a 2D fibrous structure in which interconnected nanofibrous struts form stretch-dominated polygonal cells. The historically high tensile strength of 900 MPa at a ductility of 26% renders this new type of nanofilm desirable for many technologically important applications. Apart from the flexion sensor application demonstrated in this paper, the film could be used in sweat sensors, separators for ultrathin flexible batteries, membrane distillation devices for water desalination, multiwavelength anti-reflection films, self-cleaning film coating materials, and new high-capacity capacitors. The analogous tensile property of the film to human cortical bones also suggests a potential use in the repair of fractured bone. The graphene-coated film could simultaneously act as a sensor to detect the repair progress. Besides, this breathable film may be used to save the lives of people suffering from large-area burns. The full conformability, hydrophobicity, and breathability of this film make it ideal for such applications, as it could block bacterial entry while maintaining the moisture levels required for skin regeneration and recovery.

The application of these films could also lead to significant fundamental breakthroughs. Although significant efforts have been directed to the exploration of confinement dynamics at small scales, the lack of free-standing nanostructures has led to mixed reports in the literature. We note that we have observed confinement-induced Turing patterns when casting dilute polymer solutions into the pores of the newly prepared ultrathin film *(61)*.

**ACKNOWLEDGMENTS**

The authors, particularly Professor Ping Gao, would like to express their sincere thanks to Professor Malcolm R. Mackey of the department of Chemical Engineering and Biological Technology, Cambridge University, UK. Professor Mackey provided extensive assistance with the writing of the manuscript, as well as insightful comments regarding our newly invented film. Additionally, he has promoted the film to other researchers at Cambridge University and several Australian universities after witnessing our new discovery during his visit to our laboratory in November 2017. **Funding**: The authors would also like to acknowledge the financial support received from the Research Grant Council of Hong Kong (RGC; grant number, GRF 16208215), the theme-based financial support provided by the Hong Kong Government (T23-601/17-R), and the financial support provided by the Quantum Materials Institute of HKUST. **Author contributions**: P.G. conceived the idea, R.L. and J.L. performed the experiments, Q.G. and Q.Z. helped with some experiments, M.S., T.Y., T.Z. and N.W. discussed the results, P.G. wrote the paper. **Competing interests**: The authors declare that they have no competing interests. **Data and materials availability**: All data needed to evaluate the conclusions in the paper are present in the paper and/or the Supplementary Materials. Additional data related to this paper may be requested from the authors.

**SUPPLEMENTARY MATERIALS**

**Figure S1** Photographs captured from the Movie Episode II displaying wrinkle evolution after acetone impingement at different time intervals. (A) t = 0.00 + second; (B) t = 0.04 seconds; (C) t = 0.08 seconds; (D) t = 0.12 seconds.

**Figure S2** Still images captured from **SI Movie Episode III** and **Episode IV**. (A) Optical interference pattern on the PE film at 5g indentation load. (B) Optical interference pattern at 5g load during unloading. (C) Interferences on the ruptured film.

**Figure S3** (A) A photograph of the PE film sample with inscribed lines and corresponding numerals indicating the area selected for UV-Vis spectrometry measurements. (B) Plots of UV absorption versus film thickness as a function of incident irradiation wavelength. The black circles were thickness measured using stylus profilometry. The red-, green-, and blue-lines and scattered circles represent the Beer-Lambert Law fitting as well as the calculated data points.

**Figure S4** SEM micrographs of the UHMWPE thin film surfaces. (A) Magnification ×2500; (B) Magnification ×20000.

**Figure S5** TEM image and the image analysis. (A) TEM of the PE film exhibited in Figure 2D. Arrow 1 was along the shish-kebab fibril. Arrow 3 was normal to a single but wide shish fibril. Arrow 2 was normal to two narrow shish fibrils. (B) Estimated height versus distance in the direction of the arrows measured on arrows 1 and 2. The average height of the shish fibril without the kebab is about 10 nm, and the section containing the kebab is about 19 nm. The two thin fibrils indicated by the two red peaks are about 9 nm. (C) Estimated height versus distance in the direction of arrow for position 3. The average peak height on the fibril surface is about 11 nm.

**Figure S6** BET isotherms and BJH pore size and size distributions of PE film measured at 77.3 K using nitrogen as the absorbent. (A) BET adsorption (red) and desorption (blue) isotherms; (B) BJH Pore size distributions. Left axis: derivative of absorption volume against pore diameter (gray); right axis: derivative of the absorption area against the pore diameter (light red).

**SI Movie**

**Episode I** Water jets impingement on both sides of UHMWPE ultrathin film.

**Episode II** Acetone jets impingement on one side of UHMWPE ultrathin film.

**Episode III** Load-deflection loop test (6 mm) on UHMPWE ultrathin film.

**Episode IV** Load-deflection rupture test on UHMWPE ultrathin film.

**Episode V** Dynamic Manipulation of fibrous UHMWPE film by 200 kV electron beam irradiation.

**Episode VI** Piezo-resistive flex sensor flexion and extension on wrist.

**Episode VII** Uninstallation of piezo-resistive flex sensor after flexion and extension on wrist.

Supplementary Materials for

# Flexible Ultrastrong 100-nm Polyethylene Membranes with Polygonal Pore Structures

*Jin LI*[1], Runlai LI *[1], Qiao GU[1], Qinghua ZHANG[1], T.X. YU[2], and Ping GAO[1]***

[1]Department of Chemical and Biological Engineering, [2]Department of Mechanical and Aerospace Engineering, The Hong Kong University of Science and Technology, Clear Water Bay, Kowloon, Hong Kong SAR, PRC

This file includes:

1. Characterization methods
2. Supplement data and movie: **Figures S1** to **S6**

1. MATERIALS AND METHODS

1. Characterization Methods

**Stylus profilometry**

We determined the film thickness using surface profiling measurements on the Ambios XP-2 stylus profilometer. All measurements were conducted using a tip force of 0.05 mg and the radius of the tip is 2.5 µm. To ensure the accuracy of measurements, we prepared film samples by a stepwise stacking method in which triple-, double- and single- layer PE films were laminated on a clean silicon wafer surface, see the insert in **Figure 1B** for the schematic of the stacking configuration. Specifically, we first placed cleaned silicon wafer to the film surface, and established Hamaker force between the wafer and the PE film by injecting ethanol to the wafer/polymer film interface. The Hamaker force was established by the displacement of air with permeated ethanol at the wafer/polymer or polymer/polymer interface.

**Indentation-deflection Tests**

Square-shaped film samples constrained inside carbon fiber frames with exposed area of 65 mm × 65 mm were used for the indentation-deflection test. A 3D printed cylindrical lower fixture and an upper shaft fixture with a friction-free glass ball of 15.7-mm diameter tip were used in conjunction with an ARES-2000 rheometer. Two different looped tests and one rupture test were conducted, and the repeated looped tests were conducted on two different samples. All measurements were conducted at a constant indentation speed of 25 mm/min under ambient conditions. At the end of each loop test, the sample was allowed to relax for 10 minutes before proceeding to the next loop or rupture test.

**Ultraviolet-Visible Spectroscopy**

We conducted ultraviolet-visible (UV-Vis) spectroscopy analysis on a Perkin Lambda 20 UV-vis spectrometer. We employed incident irradiation emissions with radiation areas of 1.0 mm × 9.0 mm with wavelengths between 200 nm and 1100 nm at a standard scanning rate of 480 nm/min. Prior to each measurement, we calibrated the spectrometric transmittance through air as the baseline reference.

**Scanning electron microscope (SEM)**

We conducted high-resolution scanning electron microscopic (SEM) characterizations on field emission SEM systems JEOL 7800F and JEOL 7100F. For the SEM surface characterization, we coated pure UHMWPE film onto a conductive silicon wafer surface covered with a 3-nm platinum film. The graphene/UHMWPE composite films were observed directly on SEM stubs without any prior conductive coating. To characterize the film thickness, we sandwiched the PE film between two Nafion films with approximate thicknesses of 1 μm and then fractured the composite film in liquid nitrogen. Details of the Nafion loading protocol is given in a separate manuscript that describes the use of the composite membranes as ion-exchange membranes for fuel cells and redox flow batteries (*5*). (Unpublished result). The cross-section of the fractured surface was sputter-coated with a ~4 nm-thick platinum film to minimize the charging effect.

**Transmission electron microscope (TEM)**

JEOL 2010 was used for TEM characterization, with lattice resolution of 0.19 nm. The samples used for TEM observations were prepared by directly laminating of PE films onto perforated bare TEM copper grids with individual grid dimensions of 40 × 40 μm². The van der Waals (vdW) interaction between the film and the copper grid allowed a smooth and simple attachment of the film to the copper grid.

**BET**

Brunauer–Emmett–Teller (BET) measurements were performed on the Autosorb-1, Quantachrome Instruments to collect adsorption/desorption isotherm data, with bath temperature of 77.3 K (liquid Nitrogen). Nitrogen was used as the absorbent media, with molecular weight of 28.013 g mol$^{-1}$, cross-section area of 16.2 Å$^2$, and liquid density of 0.808 g cc$^{-1}$. A total of 8.1 mg sample films were outgased at 100 ºC for 2 hours prior to analysis. The total analysis time was 820.0 minutes for both adsorption (40 pts) and desorption (40 pts).

**Tensile tests**

Uniaxial tensile property tests were carried out on the ARES-2000 rheometer using the film fixture under ambient conditions. The samples used were of dimensions of 5 mm wide × 8 mm long × ~100 nm thick. The ultimate tensile stress-strain tests were conducted at a constant Hencky strain rate of 0.1% s$^{-1}$. Six samples were tested, one each along the first and the second draw direction of the film, and four at 45º to the draw direction. Viscoelastic properties of the films were assessed by conducting loop tests at the limiting Hencky strains of 2% at Hencky strain rates of 0.1% s$^{-1}$ and 0.5% s$^{-1}$, respectively, under ambient conditions. Five loops were tested and the film was relaxed for 10 minutes at the end of each completed loop in between the loop tests.

**Mechano-Electrical Transducer Test**

Quantitative mechano-electrical responses in the above graphene coated UHMWPE film was tested in conjunction with the ARES-2000 rheometer. Film samples with the area dimension of 10 mm × 15 mm affixed with copper electrodes, wires and multimeter as described before were tested. The two copper electrodes were spaced at 4.5 mm apart and aligned parallel to the uniaxial extension/compression direction of ARES. Mechanical loop test at Hencky strain rates of 0.1% s$^{-1}$ and limiting 5% strain was performed at the ambient conditions on the film sample. Simultaneous stress-strain and ohmmic resistance-strain responses were measured.

## 2. Supplementary Figures and Movie

Video Recording at HKUST's Outdoor Sportsground

We chose to demonstrate the macroscopic characters of our newly prepared PE film by subjecting the film to alternate water/acetone jet spattering responses and conducted the experiments at HKUST's outdoor sportsground. The recordings are illustrated in **SI Movie Episode I** (water) and **II** (acetone), respectively. The DI water was produced by Millipore Milli-Q Ultra-filtration Plus water purification system. The acetone was purchased from Scharlau Company with purity of 99.5%. Both DI water and acetone were stored in 500 mL Nalgene squeeze wash bottles (LDPE) with narrow mouth nozzles. The film tested was of the dimension of 6.5 cm × 6.5 cm and an image of the film is exhibited in **Figure 1C**. Briefly, we first spattered DI water jets to both sides of the film surfaces, and then spattered acetone jets from the backside of the film only. The video recording was conducted at 25 frames per seconds (fps), and thus each frame was spaced at 0.04 seconds apart.

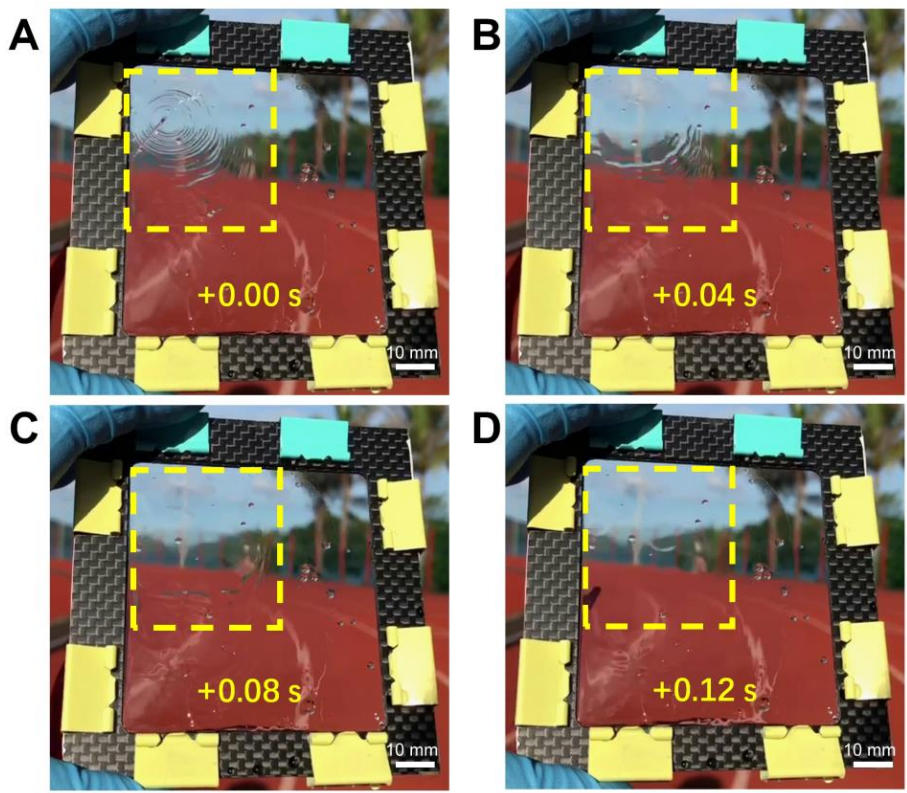

**Figure S1** Photographs captured from the Movie Episode II displaying wrinkle evolution after acetone impingement at different time intervals. (A) t = 0.00 + second; (B) t = 0.04 seconds; (C) t = 0.08 seconds; (D) t = 0.12 seconds

To illustrate the robustness and self-repair property of the film, we captured four still images from **Episode II** after an acetone jet impacted the film on the upper left corner of the film. The images are exhibited in **Figure S1**. The dotted squares inscribed in the images of **Figures S1A-D** indicate the location of acetone impingement from the back of the film. Clearly, most of the circular wrinkles generated upon impact disappeared in the first 0.04 seconds **(Figure S1B)**. Film smoothness was fully restored after 0.12 seconds of acetone jet impact **(Figure S1D)**.

Video Recording of the Indentation-Deflection Responses

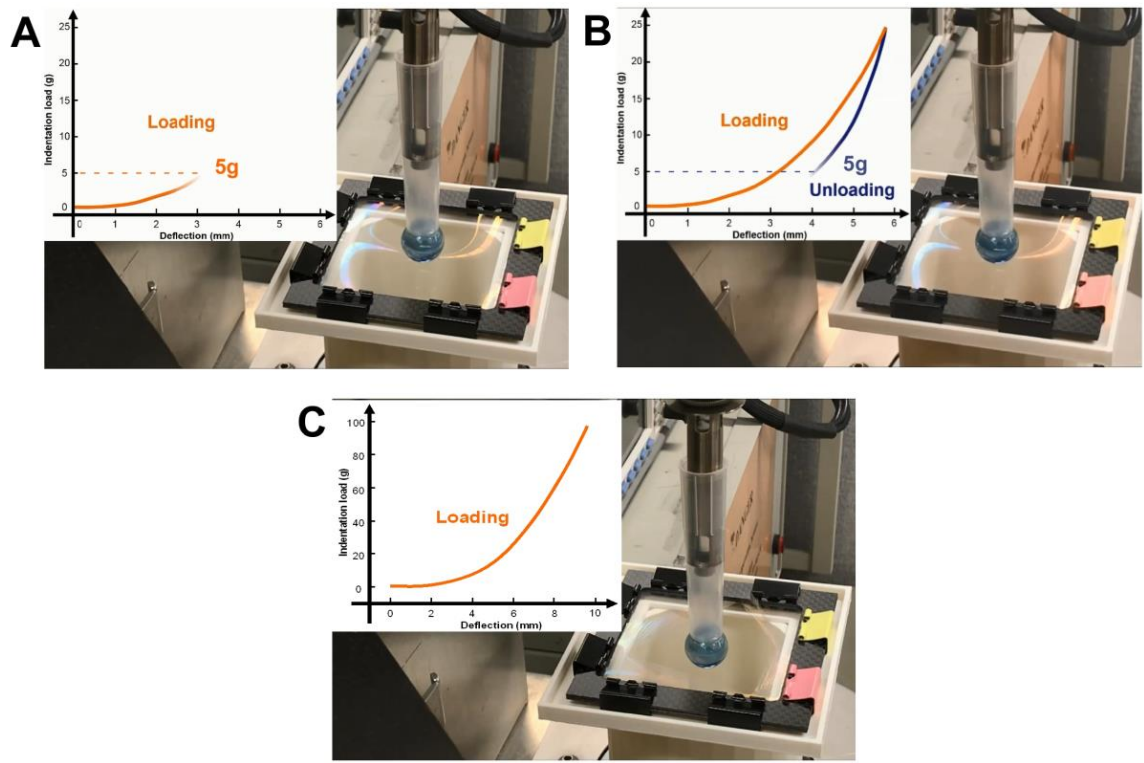

**Figure S2** Still images captured from **SI Movie Episode III** and **Episode IV**. (A) Optical interference pattern on the PE film at 5g indentation load. (B) Optical interference pattern at 5g load during unloading. (C) Interferences on the ruptured film.

To quantitatively characterize the robustness and flexibility of the PE film, we conducted indentation-deflection measurements on a customized indentation setup. Again, the specimen was of the dimension of 65 mm × 65 mm as described above. The sample was subjected to loop tests at limiting deflections of 6 mm, 8 mm and finally to rupture. All tests were conducted at a constant speed of 25 mm/min. Video recordings of real time load-displacement responses as well as the mechano-optical interferences are shown in **SI Movie Episodes III-IV**.

The test protocol was as follows. Firstly, the film was loaded onto a 3D printed ABS fixture, which was attached to the lower fixture under the ARES-2000 rheometer. Secondly, a friction-free 15.7-mm-diameter glass ball was affixed to a tubular shaft connected to the upper fixture of the ARES-2000 rheometer. Thirdly, we adjusted the setup such that two parallel graded interference patterns (yellowish red strips) formed between the film and the ceiling lights are captured close to the center of the PE film surface. Fourthly, load-displacement loops at 6-mm limiting deflection were imposed on the sample at 25 mm/min. Upon completion of the full unloading test, the sample was allowed to relax for 10 minutes, and a second loop test at 8-mm limiting deflection was imposed on the sample and the sample was allowed to relax for 10 minutes at the cessation of the loop test. Finally, ultimate rupture test was carried out on the sample.

To illustrate how the film might be a potential mechano-optical sensor, we captured two still images from **Episode III** at 5 gf load during the loading **(Figure S2A)** and unloading (Figure **S2B**) process of the loop test. The film's high toughness and thinness is depicted in the image captured upon rupture and exhibited in **Figure S2C**.

UV-Vis Spectroscopy and Film Thickness Measurements

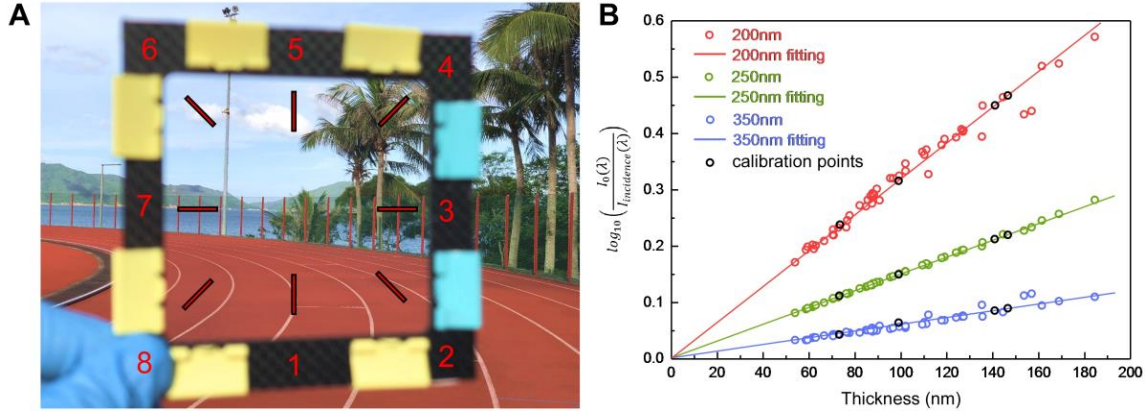

**Figure S3** (A) A photograph of the PE film sample with inscribed lines and corresponding numerals indicating the area selected for UV-Vis spectrometry measurements. (B) Plots of UV absorption versus film thickness as a function of incident irradiation wavelength. The black circles were thickness measured using stylus profilometry. The red-, green-, and blue-lines and scattered circles represent the Beer-Lambert Law fitting as well as the calculated data points.

According to Lambert-Beer law, at certain wavelength, the thickness can be calculated based on the absorption, $A$, of an incident light passing through the PE film:

$$A = log_{10}\left(\frac{I_0(\lambda)}{I_{incidence}(\lambda)}\right) = \varepsilon(\lambda)lc$$

where $I_0$ and $I_{incidence}$ are the intensity of the transmittance of the irradiated light through air and the film sample, respectively. $\varepsilon(\lambda)$ is the absorptivity coefficient that depends on the wavelength of the incident light, $c$ is concentration, and $l$ is the pathlength, or the thickness of the membrane if the light is irradiated perpendicular to the plane of the film surface.

Thus, if the sample is of similar structure such that $c$ and $\varepsilon(\lambda)$ are constant, we can use the equation to estimate the film thickness by measuring the absorption intensity $A$. To validify this simple method for thickness measurements, we first correlated the absorption intensity with film thickness using stylus profilometry. The calibration and measurement protocols are as follows. First, UV-vis spectroscopic tests were conducted on a film sample at eight different locations on the film, see **Figure S3A**. Secondly, four different films with different initial thickness and measured absorption intensity were then transferred to cleaned wafer surfaces for stylus profilometry. Using the average film thickness measured and the absorption intensity, we obtained the product $\varepsilon(\lambda)c$ in the Lambert-Beer equation. The tested sample points are exhibited as black circles in **Figure S3B.** These values are then further used for the prediction of film thicknesses at different absorption intensities at 200 nm, 250 nm and 300 nm, respectively. The correlation R for these wavelengths is 0.9876, 0.9996 and 0.9711, for 200 nm, 250 nm and 350 nm, respectively. The nearly perfect linearity, particularly at 250 nm incidence irradiation demonstrates that UV-Vis spectroscopy is a facile and non-invasive measurements for determination of film thickness.

SEM of the Film Surface

The lower magnification images in **Figure S4A** and **S4B** show the homogeneity of the film. The inscribed dotted yellow rectangle with the dimension of 2.4 μm × 1.8 μm in the middle of SEM micrographs depicts the position of the micrograph exhibited in **Figure 2B**

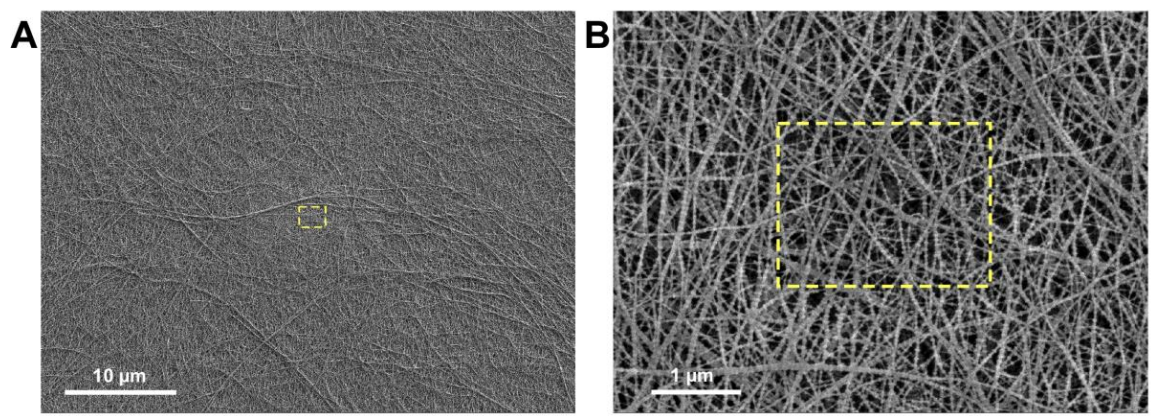

**Figure S4** SEM micrographs of the UHMWPE thin film surfaces. (A) Magnification ×2500; (B) Magnification ×20000.

TEM Image Analysis

The TEM in **Figure S5A** is the same as that in **Figure 2D** except it contains three inscribed arrows. We performed image analysis along these arrows to show that the electron scattering intensities are independent of fibril widths. The analysis data exhibited in **Figure S5B and S5C** were obtained using Gwyddion 2.50 (free-SPM data analysis software), the "extract profiles" function to interpret the grey scale into the thickness difference (*6*). The analysis procedure is as follows. First, we imported the TEM micrograph into gwyddion and define its 3D dimensions as 440 nm × 440 nm × 100 nm. Second, we performed a three-points leveling (20-nm-diameter for each point) on the imported TEM image. We preassumed that the gray-scale contrast of pores at different locations was the same. Third, we extracted the profiles along the lines drawn at positions shown in **Figure S5A**, the widths of these lines were 10 nm. Then the thickness could be directly obtained as shown in **Figure S5B** & **C**.

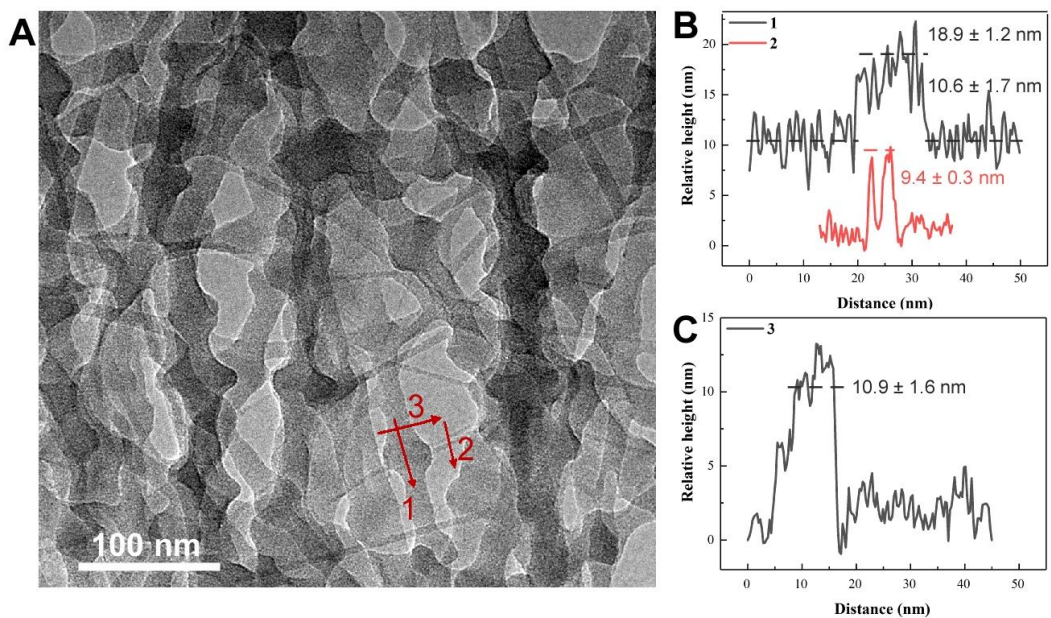

**Figure S5** TEM image and the image analysis. (A) TEM of the PE film exhibited in Figure 2D. Arrow 1 was along the shish-kebab fibril. Arrow 3 was normal to a single but wide shish fibril. Arrow 2 was normal to two narrow shish fibrils. (B) Estimated height versus distance in the direction of the arrows measured on arrows 1 and 2. The average height of the shish fibril without the kebab is about 10 nm, and the section containing the kebab is about 19 nm. The two thin fibrils indicated by the two red peaks are about 9 nm. (C) Estimated height versus distance in the direction of arrow for position 3. The average peak height on the fibril surface is about 11 nm.

**SI Movie, Episode V** presents *in situ* video recordings of the PE film edges on the copper grid under manipulation by charges produced by electron beam irradiation at an accelerating voltage of 200 kV. The fibrous UHMWPE film is shown to flip up and down, like the mechanical opening and closing of a book. This experiment further demonstrates the thinness and porosity of the new UHMWPE film.

BET Characterization

**Figures S6** shows the BET absorption isotherms of the PE film. The absorption isotherm displayed in **Figure S6A** is a type IV isotherm, and from the nearly constant absorption isotherm volume at relative pressures up to 0.9 and the corresponding BJH pore size distribution data in **Figure S6B**, we can conclude that the pores in the PE film are flat pores.

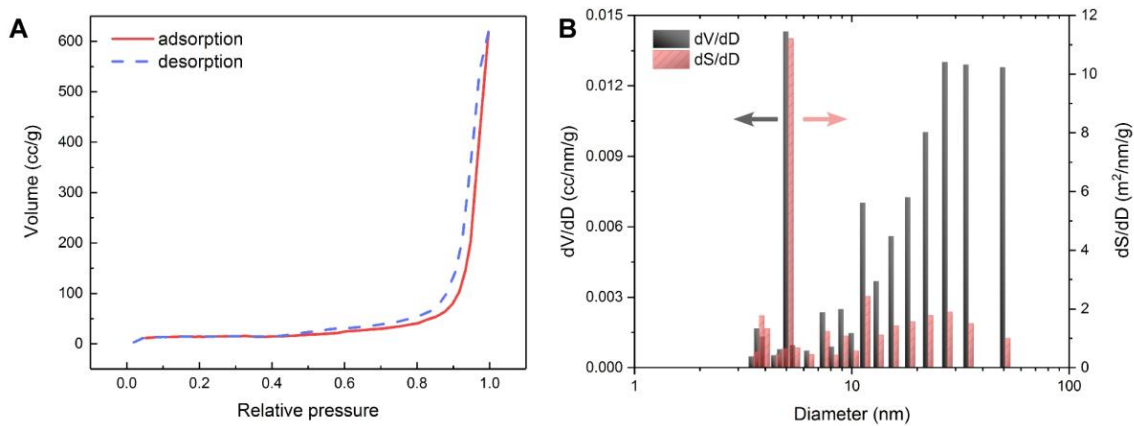

**Figure S6** BET isotherms and BJH pore size and size distributions of PE film measured at 77.3 K using nitrogen as the absorbent. (A) BET adsorption (red) and desorption (blue) isotherms; (B) BJH Pore size distributions. Left axis: derivative of absorption volume against pore diameter (gray); right axis: derivative of the absorption area against the pore diameter (light red).